# Carbon nanotube-sharp tips and carbon nanotube-soldering irons

A. Misra[1], C. Daraio[1,2]

[1] Graduate Aeronautical Laboratories (GALCIT),

[2] Applied Physics

California Institute of Technology, Pasadena, CA, 91125

**We report on the nano-electron beam assisted fabrication of atomically sharp iron-based tips and on the creation of a nano-soldering iron for nano-interconnects using Fe-filled multiwalled carbon nanotubes (MWCNTs). High energy electron beam machining has been proven a powerful tool to modify desired nanostructures for technological applications (*1-4*) and to form molecular junctions and interconnections between carbon nanotubes (*2, 5*). Recent studies (*5*) showed the high degree of complexity in the creation of direct interconnections between multiwalled and CNTs having dissimilar diameters. Our technique allows for carving a MWCNT into a nanosoldering iron that was demonstrated capable of joining two separated halves of a tube. This approach could easily be extended to the interconnection of two largely dissimilar CNTs, between a CNT and a nanowire or between two nanowires.**

The development of the next generation of miniaturized electronic systems demands the integration of nanoelectronic components creating reliable mechanical and electrical contacts. Despite many years of carbon nanotubes research, the electrical interconnection of individual nanotubes and nanowires remains a tedious task. At the same time, the development of scanning probe techniques and magnetic recording media require an ever decreasing tip size of ultrasharp magnetic read-write heads (*6*). Recently, focused electron beams have been reported for the machining of both single walled (SWCNTs) and multiwalled (MWCNTs) nanotubes, for cutting them or for forming junctions at the atomic level (the latter being limited so far to only small and similar diameter tubes) (*5*). Along with the well known electrical, mechanical and thermal properties of the carbon nanotubes, the presence of a metal filling inside their core has been suggested to enhance their applicability in magnetic recording media, new electronic devices and to reinforce the material durability (*7,8*). Here we report two interesting phenomena of nanoelectron beam assisted machining of iron-filled multiwalled carbon nanotubes (MWCNTs): the fabrication of atomically sharp magnetic tips and the direct interconnection of two large diameter MWCNTs. Previous attempts to solder CNTs had been performed through metal (*5*) and amorphous carbon (*2*) deposition at the tips, but the results showed little or questionable success.

The Fe-filled carbon nanotubes (CNTs) considered for this study were produced by dissolution of a catalyst source and a carbon source in a two-stage thermal chemical vapor deposition system. This system consists of a 30 mm diameter, and 1000 mm long quartz tube inserted in a tube furnace having 200 mm preheating (at 80 $^0$C) and 500 mm heating (at 825 $^0$C) zone. Si was used as growth substrate while a mixture of Fe-catalyst (ferrocene) and carbon source (toluene) (0.02 g/ml) were injected into the preheating zone at the rate of 5 ml/15 min. A flow of 100 SCCM of argon gas was maintained as carrier for the solution into the heating zone. The grown MWCNTs had outer diameters ranging from 20-100 nm and their length was ~100 μm. For the transmission electron microscope analysis and machining, a small amount of the grown material was scratched from the substrate, dispersed in iso-propyl alcohol and deposited on a holey carbon coated TEM grid. Electron irradiation for manipulation, etching and imaging was obtained in a TEM (FEI Technai F-30 UT) with a field emission gun operating at an acceleration voltage 300 kV at room temperature (with no heating stage attached to the specimen holder). The field emission gun together with the probe forming lenses of the microscope is capable of producing a nanometer-sized electron beam with current density of the order $\sim 1.3 \times 10^3$ A/cm$^2$. Irradiation was carried out at beam current densities varying from $1.3 \times 10^3$ A/cm$^2$ to $7.5 \times 10^2$ A/cm$^2$ on different diameter multiwalled carbon nanotubes. Sequences of TEM images were recorded using ORCA-ER camera with 1280 x 1024 pixel format in a fixed bottom mount configuration and this camera uses a Hamamatsu DCAM supported board for acquisition.

We first describe the creation of atomically sharp magnetic tips (a schematic diagram of the processing and resultant structures is reported in Fig. 1A-D) and later the synthesis and operation of a CNT-soldering iron (Fig. 1E-F). It was shown that under controlled irradiation with a highly focused electron beam, Fe-filled MWCNTs can be cut at selected locations (*9*). Using similar experimental conditions, we slice a MWCNT in two parts (Fig. 1B) and expose the Fe nanorods encapsulated inside the core of the tube to a continuous e-beam, centered in close proximity of the cut. This high energy irradiation causes an increase of the tube's internal pressure, localized joule heating and related electro-migration (*5*), that lead to a modification of the Fe nanorods's properties (*10*), and to surface reconstruction (*11*) at the incised rim. We report TEM based observations of the described steps for the synthesis of an atomically sharp Fe tip in Fig. 2. The MWCNT studied had inner and outer diameters of 9 nm and 32 nm respectively (Fig. 2A). Electron beam cutting was achieved dragging a nano e-beam (spot size 8 nm in diameter) with a current density of $1.3 \times 10^3$ A/cm$^2$ along the CNT diameter. Figure 2B shows the process of formation of the opening in the MWCNT along the focused e-beam path. After few seconds of the cutting, the surface reconstruction at the unstable open ends of the carbon nanotube walls as well as on the surface of the Fe-nanorod was evident, as shown in Fig 2 B and C (and pointed by the arrow). It is clear that due to a strong tendency towards the reduction of the surface energy (*12*), the capping of the open wall ends occurs with the formation of stable closed structure by cross linking between adjacent graphene planes (inter-walls interaction) and the generation/migration of vacancy-interstitials pairs (*13*) due to the structural damage. The dangling atoms at the edge of the cut combined with the localized high temperature may cause the formation of carbon dimer $C_2$ units (*14, 15*) which after recombination form pentagonal rings, the most essential building blocks in the formation of curved structures. This phenomenon results in the formation of semi-fullerene-like caps interlinking the walls and healing the incised surface. As a result the Fe nanorod near the incision remains trapped between the graphene caps on one side and the CNT's walls on the other. The self compression and dynamic behavior of encapsulated material inside graphitic networks such as carbon onion or multiwalled carbon nanotubes was the subject of recent investigations (*10, 16*). It was shown that under electron beam irradiation locally high internal pressure and temperature caused phase modification and shape changes of the metallic particles trapped inside the core. Upon continuous electron irradiation, the damage and reconstruction of the outermost carbon lattice further increased the pressure within the graphene cells, enhancing such effects. Similarly, in our experiments it is evident that immediately after the reconstruction of the sliced walls' edges into a closed cell structure (Fig. 2A), the CNT starts behaving as a high pressure cell (*10*) for the encapsulated Fe nanorod. Because of the high pressure and temperature, Fe is pushed out from the nanotube's core and local recrystallization phenomena occur. At the same time, a continuous etching of the outer carbon walls occurs, causing the incised nanotube's edge to assume a pointed pencil-like shape. To further investigate the effects of electron irradiation, we investigated the time evolution of the encapsulated Fe nanorod structure during the beam exposure of one side of the cut (upper edge of Fig. 2B). The Fe nanorod protrusion and the tip formation are shown in Fig. 2C-D after ~17 min of e-beam irradiation. In addition to the outward extrusion of the nanorod towards the pressure gradient, the e-beam interaction also induces melting and re-crystallization on the exposed metal surface. This is made possible by self diffusion processes caused by the small-size effect (*17,18*) that reduces the melting temperature of the nanorod. The higher interfacial free energy per unit area increases the tendency of melting at the surface at a temperature below the thermodynamic equilibrium melting temperature. Recently, the phenomenon of liquid-state surface faceting as a precursor to surface induced crystallization was being observed in metal–alloy system (*4*). This was considered as one of the significant hallmarks of the crystalline state, as stable facets with low specific surface free energy determine the equilibrium shape of solid particles. In our experiments we observe the equilibrium faceting of Fe after recrystallization into a hexagonal shape (Fig. 2D). This enables the formation of an atomically sharp tip whose outer diameter is ~7 nm, and the vertex size of <1 nm. The complete formation of the recrystallized probe after ~17 min of e-beam irradiation is shown in Fig. 2E. Because of the absence of a heating stage in our system, we could not preserve the initial parallel walls-tube structure, as the low defect mobility at room temperature prevents reconstruction and leads to rapid destruction of the graphite lattice. To ensure the graphite lattice reconstruction the specimen should be maintained at a temperature of >300 $^0$C (*19*).

The melting and recrystallization effects described above for the formation of the sharp tips suggest the possibility of utilizing the same structures for nano-soldering and interconnects, simply by prolonging the e-beam exposure of the tips to enable additional ejection of the melt metal. Such tips could then be used as nanosoldering irons if places in proximity of one or two other elements to be connected. To explore the effective soldering ability of such probes, we performed a systematic investigation on various Fe-filled MWCNTs of different inner and outer diameters (20 and 60 nm) by using the same condition of electron irradiation. Similar processing steps were followed for the cutting (Fig. 3A) and etching (Fig. 3B and 3C) of the tubes, with variable exposure time dependent on the different tube's diameters. It was demonstrated that the prolonged local electron irradiation on both sides of the incision causes an outward flow of the melted Fe-nanorods that "grow" towards each other until their successful final soldering. The TEM image reported in Fig. 3C shows a complete merging and "healing" of two previously separated halves of the CNT after ~30 min of e-beam exposure. From these results it is clear that time for cutting and soldering of different nanotubes and/or nanowires depends on the outer diameter and number of walls composing the nanotubes. A higher magnification image of the soldered zone is shown in Fig. 3D. The aggregation and recrystallization of Fe at the junction is noticeable, and it is evident that surface modification phenomena took place all over the nanowire's area exposed to the e-beam. Remarkably, two distinct features were observed in the soldering process between the two Fe nanowires: the outward flow of the metal due to the pressure gradient created by the e-beam energy, and its surface melting and re-crystallization forming at the soldered zone a polycrystalline junction. It would be interesting to study systematically the electrical properties of such junctions under variable irradiation condition and different soldering materials. To further characterize the interconnection-structure evolution, we recorded electron diffraction patterns (see inset of Fig. 3D) from the re-crystallized area, after completing the tube's reconnection. We found scattered low intensity spots from polycrystalline iron and the presence of diffused rings typical of disordered carbon. A low magnification image of the final stage of the soldering process is shown in Fig. 3E.

We analyzed the Fe nanorod "growth" rate in terms of volume of metal exposed outside its original CNT-enclosure as a function of irradiation time on two samples differing in inner and outer diameters (Fig. 4 A and B). The plots show in all cases a rather nonlinear behavior, likely related to two leading phenomena: e-beam etching of the CNT's wall in the initial phase, and later melting, flowing and recrystallization of the metal. From these results it is clear that the amount of time necessary for nanosoldering depends on the geometry of the samples.

Through this work we present the first experimental demonstration of the use of nano-electron beam engineering of MWCNTs as an advantageous tool for the creation of atomically sharp Fe-based tips. In addition, we report the creation of a carbon nanotube soldering iron and proved its effectiveness in the connection of two MWCNTs. The same approach can be use for soldering a variety of different nanostructures. This system represents a viable tool for interconnecting nanowires, for example in creating asymmetric heterostructures and heterojunctions (*20*). Although far from being completely characterized, this work adds a new functionality in nanoelectromechanical systems and integrated circuitry that might dramatically change the engineering of nanoelectronic devices.

**Acknowledgements** C.D. wishes to acknowledge the support of this work by Caltech start-up funds, A.M. acknowledges support by the Moore Fellowship. This work benefited from use of the Caltech KNI and Mat Sci TEM facilities supported by the MRSEC Program of the National Science Foundation under Award Number DMR-0520565. We thank Prof. Florian Banhart for helpful discussions.

**Correspondence** and requests for materials should be addressed to daraio@caltech.edu.

**Figure 1.** Processing steps for the formation of an atomically sharp Fe tip and of the soldering between two CNTs-halves under electron beam irradiation. **A**, As-grown Fe-encapsulated MWCNT. **B**, Electron-beam assisted cutting of a gap. **C-D**, Growth of the metal tip. **E-F**, Melting, flowing and final soldering back together of the CNT's halves.

**Figure 2.** TEM snapshots showing the melting, flowing and recrystallization during the formation of atomically sharp Fe tip after 300 kV electron beam irradiation. **A**, TEM image showing the pristine CNT with Fe nanorod (catalyst) encapsulated in its inner core. **B**, Cutting of a 10 nm gap across the tube by moving a highly focused electron beam. The area marked by a white circle underlines the exposure of Fe nanorod. **C**, High magnification image of the MWNTs walls reconstruction into closed fullerene-like caps (marked by the arrow). **D**, Complete formation of the sharp Fe-tip with hexagonal geometry after 17 min. **E**, MWCNT-supported sharp nanosoldering iron.

**Figure 3.** TEM images showing the nanosoldering of a cut Fe-filled MWCNT. **A**, Initial incision and gap-opening. **B**, Etching of the carbon walls surrounding the incision by e-beam irradiation (and initial exposure of the nanowire). **C**, Soldering back together of the Fe nanorods after etching, extrusion and surface re-growth. **D**, High resolution image of the soldered area showing polycrystalline Fe. Inset shows a diffraction pattern recorded in the area. **E**, Low magnification image showing the completed soldering process.

**Figure 4.** Growth rate of the Fe nanorod outside of its original CNT enclosure expressed as exposed volume ($nm^3$) as a function of time (minutes) for **A**, 9 inner/32 nm outer diameter tube and **B**, for the nanosoldering process described in Fig. 3.

**Figure.1**

a  b  c

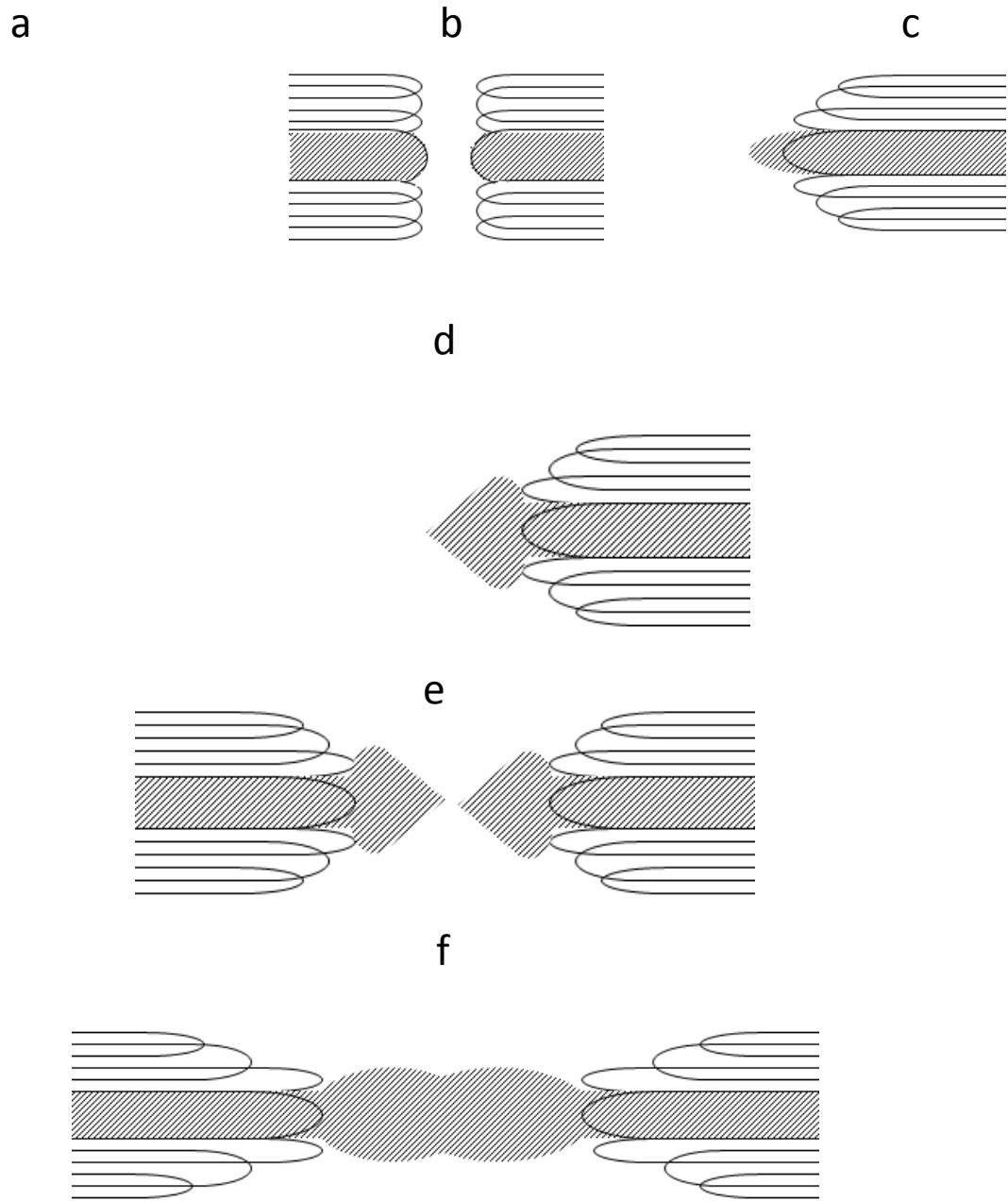

d

e

f

**Figure. 2**

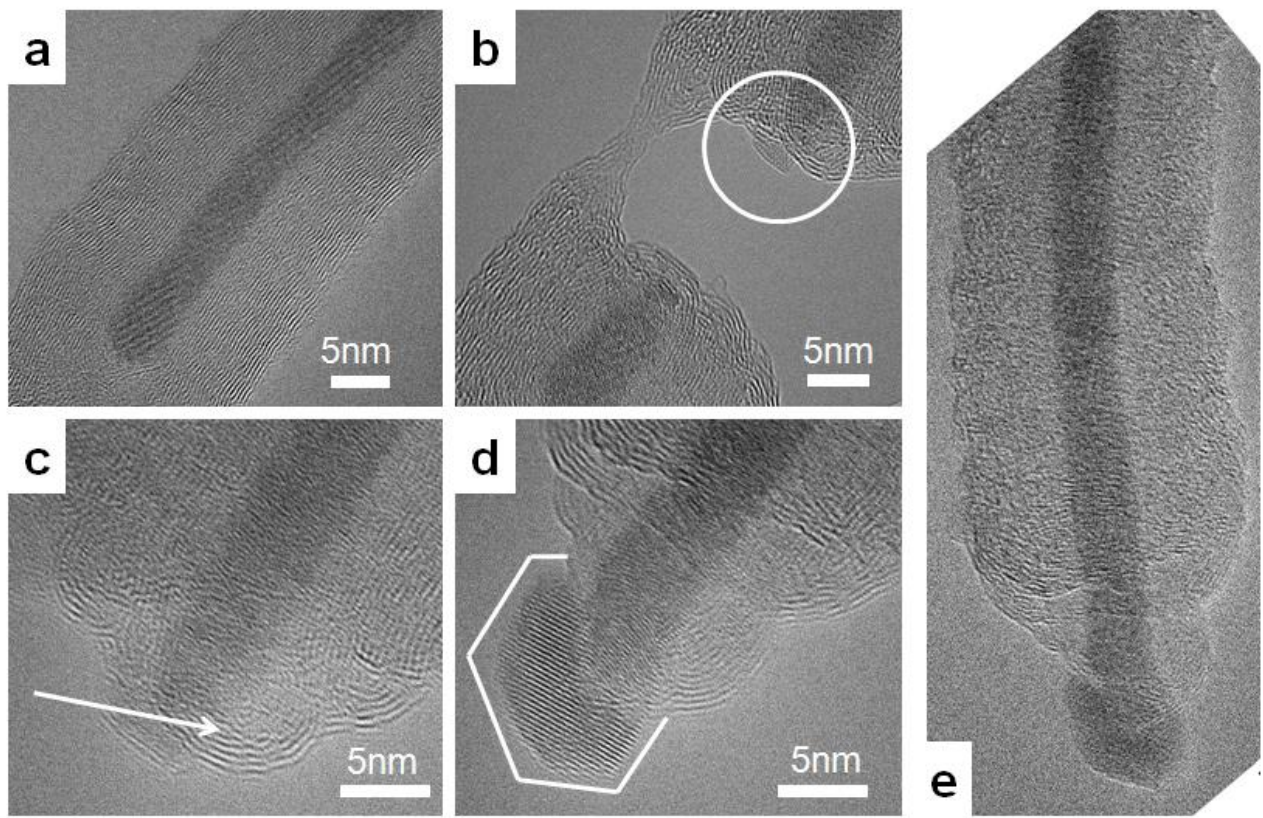

Figure.3

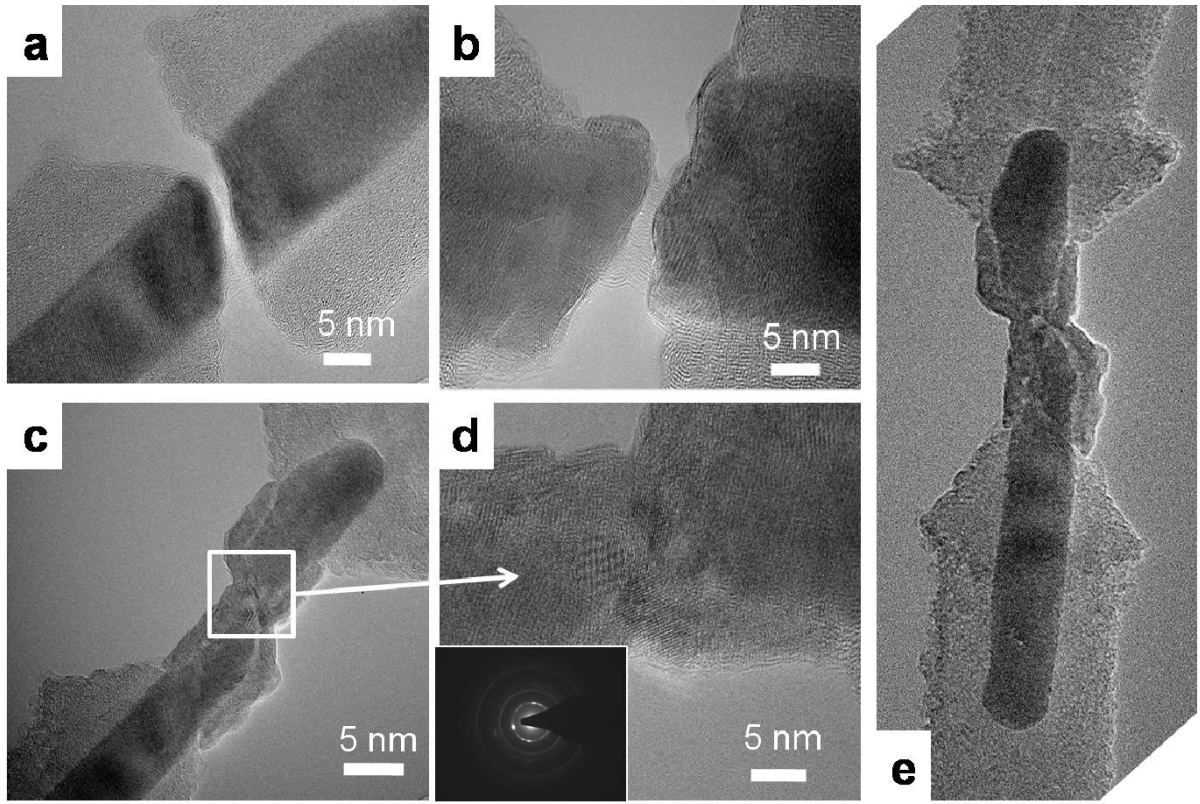

**Figure.4**

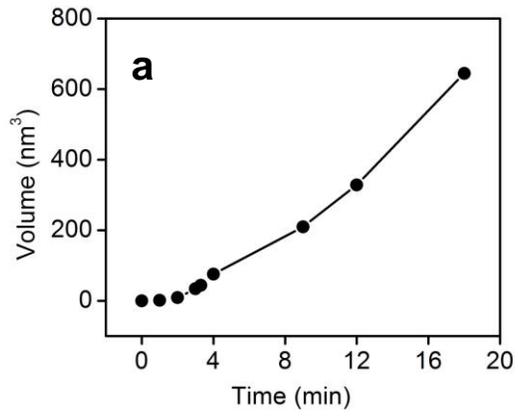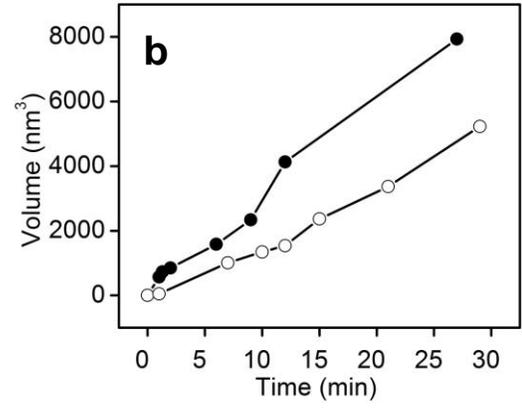